\documentstyle[12pt]{article}
\newcommand{\bea}{\begin{eqnarray}}
\newcommand{\eea}{\end{eqnarray}}
\newcommand{\be}{\begin{equation}}
\newcommand{\ee}{\end{equation}}
\newcommand{\vs}[1]{\vspace{#1 mm}}

\newcommand{\dsl}{\pa \kern-0.5em /}

\newcommand{\pa}{\partial}

\newcommand{\nn}{\nonumber\\}

\begin{document}
\topmargin 0pt
\oddsidemargin 0mm



\begin{flushright}

USTC-ICTS-09-03\\



\end{flushright}

\vspace{2mm}

\begin{center}

{\Large \bf Interaction between two non-threshold bound states

}

\vs{6}

{\large J. X. Lu\footnote{E-mail: jxlu@ustc.edu.cn}, Bo
Ning\footnote{E-mail: nwaves@mail.ustc.edu.cn}, Ran
Wei\footnote{E-mail: wei88ran@mail.ustc.edu.cn}
 and Shan-Shan Xu\footnote{E-mail: xuss@mail.ustc.edu.cn}}

 \vspace{4mm}

{\em

  Interdisciplinary Center for Theoretical Study\\

 University of Science and Technology of China, Hefei, Anhui
 230026, China\\





 }

\end{center}

\vs{15}

\begin{abstract}

A general non-threshold BPS (F, D$_p$) (or (D$_{p - 2}$, D$_p$))
bound state can be described by a boundary state with a quantized
world-volume electric (or magnetic) flux and is characterized by a
pair of integers $(m, n)$. With this, we calculate explicitly the
interaction amplitude between two such non-threshold bound states
with a separation $Y$ when each of the states is characterized by a
pair of integers ($m_i, n_i$) with $i = 1, 2$. With this result, one
can show that the non-degenerate (i.e., $m_i n_i \neq 0$)
interaction is in general attractive for the case of (D$_{p - 2}$,
D$_p$) but this is true and for certain only at large separation for
the case of (F, D$_p$). In either case, this interaction vanishes
only if $m_1/ n_1 = m_2/ n_2$ and $n_1 n_2 > 0$. We also study the
analytic structure of the corresponding amplitude and  calculate in
particular the rate of pair production of open strings in the case
of (F, D$_p$).

\end{abstract}
\newpage
\section{Introduction}

It is well-known by now that two parallel Dp-branes separated by a
distance feel no force between them, independent of their
separation, when they are both at rest. This is due to the BPS
nature or the preservation of certain number of space-time
supersymmetries of this system, and goes by the name ``no-force"
condition. This was shown initially for brane supergravity
configurations through a probe
\cite{Dabholkar:1990yf,Duff:1993ye,Duff:1994an} and later through
the string level computations as an open string one-loop annulus
diagram with one end of the open string located at one D-brane and
the other end at the other D-brane making use of the ``usual
abstruse identity" \cite{Polchinski:1995mt}. With this feature, one
can easily infer that when one of branes in the above is replaced by
the corresponding anti-brane, there must be a separation-dependent
non-vanishing force to arise since such a system is not a BPS one
and breaks all the space-time supersymmetry. The corresponding
forces can easily be computed given our knowledge of computing
forces between two identical branes. In general, no
separation-dependent force arising is a good indication that the
underlying system preserves certain number of space-time
supersymmetries.

In addition to the simple strings and simple D-branes, i.e.,
extended objects charged under only one NS-NS potential or one R-R
potential, there also exist their supersymmetry preserving bound
states such as (F, D$_p$) \cite{Witten:1995im, Schmidhuber:1996fy,
Arfaei:1997hb, Lu:1999qia, Lu:1999uca, Lu:1999uv, Hashimoto:1997vz,
DiVecchia:1999uf} and (D$_{p - 2}$, D$_p$)
\cite{Breckenridge:1996tt, Costa:1996zd, Di Vecchia:1997pr}, i.e.,
extended objects charged under more than one potential. It would be
interesting to know how to compute the forces between two such bound
states separated by a distance. Since each of the bound states
involves at least two kinds of branes, the force structure is richer
and more interesting to explore. In this paper, we will focus on the
above mentioned two types of the so-called non-threshold BPS bound
states, namely (F, D$_p$) and (D$_{p - 2}$, D$_p$), with even $p$ in
IIA and odd $p$ in IIB, respectively.

The non-threshold BPS bound state (F, D$_p$), charged under both
NS-NS 2-form potential and R-R $(p + 1)$-form potential, is formed
from the fundamental strings and D$_p$ branes by lowering the system
energy through dissolving the strings in the D$_p$ branes, turning
the strings into flux. The similar picture applies to the
non-threshold BPS (D$_{p - 2}$, D$_p$) bound state charged under
both R-R $(p - 1)$-form potential and R-R $(p + 1)$-form potential,
where the initial D$_{p - 2}$ branes dissolve in D$_p$ branes,
turning into flux, too.  Dirac charge quantization implies that the
two potentials for either bound state are characterized by their
corresponding quantized charges, therefore each bound state is
characterized by a pair of integers $(m, n)$. When the pair of
integers is co-prime, the system is stable (otherwise it is
marginally unstable) \cite{Schwarz:1995dk}. In this paper, we will
use the description of a boundary state with a quantized
world-volume flux given in \cite{Di Vecchia:1997pr,DiVecchia:1999uf,
Di Vecchia:1999fx} for the bound state to calculate explicitly the
interaction between two non-threshold (F, D$_p$) (or (D$_{p - 2}$,
D$_p$)) bound states separated by a distance. Here each state is
characterized by an arbitrary pair of integers $(m_i, n_i)$ with $i
= 1, 2$. We find that the non-degenerate (i.e., $m_i n_i \neq 0$)
force is in general attractive for the case of (D$_{p - 2}$, D$_p$)
but this is only certain at large separation for the case of (F,
D$_p$). This interaction in either case vanishes only if $m_1/n_1 =
m_2/n_2$ and $n_1 n_2 > 0$. The expected vanishing interaction for
the special case of two identical (F, D$_p$) bound states was
previously shown in \cite{DiVecchia:1999uf}.

This paper is organized as follows. In the following section, we
will review the boundary state with a given external field,
therefore providing a possible representation for the non-threshold
(F, D$_p$) or (D$_{p - 2}$, D$_p$) bound state. In addition, we
present the various couplings of the boundary state to bulk massless
fields and set the conventions for the following sections. In
section 3, we  calculate the long-range interaction between two (F,
D$_p$) (or (D$_{p - 2}$, D$_p$)) bound states separated by a
distance $Y$ with each state characterized by an arbitrary pair of
integers $(m_i, n_i)$ ($i= 1, 2$), and study the underlying
properties. In section 4, we calculate the interaction at the string
level between two arbitrary (F, D$_p$) (or (D$_{p - 2}$, D$_p$))
bound states placed parallel to each other with a separation $Y$
using the closed string boundary state approach. We summarize the
results in section 5.

\section{The boundary state and its couplings}

We in this section briefly review what we need about the boundary
state of D-branes with a constant external field on the world-volume
as well as its couplings to various bulk massless modes. In
addition, we present the derivation of these couplings through the
D-brane effective action with a constant world-volume field and set
the conventions for this paper. The material of this section is
largely taken from \cite{Billo:1998vr,Di Vecchia:1997pr,
DiVecchia:1999uf, Di Vecchia:1999rh, Di Vecchia:1999fx} and the
detail is referred to those papers.

\subsection{The boundary state with an external world-volume field}

In the closed string operator formalism, the supersymmetric BPS
D-branes of type II theories can be described by means of boundary
states $|B\rangle$ \cite{Callan:1986bc,Polchinski:1987tu}. For such
a description, we have two sectors, namely NS-NS and R-R sectors,
respectively. Both in the NS-NS and in R-R sectors, there are two
possible implementations for the boundary conditions of a D-brane
which correspond to two boundary states $|B, \eta\rangle$ with $\eta
= \pm$. However, only the following combinations \be |B\rangle_{\rm
NS} = \frac{1}{2}\left[ |B, +\rangle_{\rm NS} - |B, -\rangle_{\rm
NS} \right],\ee and \be |B\rangle_{\rm R} = \frac{1}{2} \left[ |B,
+\rangle_{\rm R} + |B, -\rangle_{\rm R} \right]\ee are selected by
the GSO projection in the NS-NS and in the R-R sectors,
respectively. The boundary state $|B, \eta\rangle$ is the product of
a matter part and a ghost part \cite{Billo:1998vr} as \be |B,
\eta\rangle = \frac{c_p}{2} |B_{\rm mat}, \eta\rangle |B_{\rm g},
\eta\rangle,\ee where \be |B_{\rm mat}, \eta\rangle = |B_X \rangle
|B_\psi, \eta\rangle, \qquad |B_{\rm g}, \eta\rangle =
|B_{gh}\rangle |B_{sgh}, \eta\rangle .\ee The overall normalization
$c_p$ can be unambiguously fixed from the factorization of
amplitudes of closed strings emitted from a disk
\cite{Frau:1997mq,Di Vecchia:1997pr} and is given by \be c_p =
\sqrt{\pi}\left(2\pi \sqrt{\alpha'}\right)^{3 - p} .\ee The explicit
expressions of the various components of $|B\rangle$ as indicated
above are given in \cite{Billo:1998vr} in the case of a static
D-brane without any external field on its world-volume. However, as
discussed in \cite{DiVecchia:1999uf}, the operator structure of the
boundary state does not change even when more general configurations
such as the presence of an external field on the world-volume are
considered and is always of the form \be |B_X\rangle = {\rm
exp}\left[-\sum_{n =1}^\infty \frac{1}{n} \alpha_{- n} \cdot S \cdot
{\tilde \alpha}_{ - n}\right] |B_X\rangle^{(0)},\ee and \be |B_\psi,
\eta\rangle_{\rm NS} = - {\rm i}~ {\rm exp} \left[{\rm i}\, \eta
\sum_{m = 1/2}^\infty \psi_{- m} \cdot S \cdot {\tilde \psi}_{- m}
\right] |0\rangle \ee for the NS-NS sector and \be |B_\psi,
\eta\rangle_{\rm R} = - {\rm exp} \left[{\rm i}\, \eta \sum_{m =
1}^\infty \psi_{- m} \cdot S \cdot {\tilde \psi}_{- m} \right] |B,
\eta\rangle_{\rm R}^{(0)} \ee for the R-R sector\footnote{The phases
chosen in (7) and (8) are just for the convenience when we compute
the couplings to various bulk massless modes. }. The matrix $S$ and
the zero-mode contributions $|B_X\rangle^{(0)}$ and $|B,
\eta\rangle_{\rm R}^{(0)}$ encode all information about the overlap
equations that the string coordinates have to satisfy, which in turn
depend on the boundary conditions of the open strings ending on the
D-brane. Since the ghost and super-ghost fields are not affected by
the type of the boundary conditions imposed, the corresponding part
of the boundary state remains the same and its explicit expressions
can be found in \cite{Billo:1998vr}. We would like to point out that
the boundary state must be written in the $(-1, -1)$ super-ghost
picture in the NS-NS sector, and in the asymmetric $(- 1/2, - 3/2)$
picture in the R-R sector in order to saturate the super-ghost
number anomaly of the disk \cite{Friedan:1986,Billo:1998vr}.

Given what has been said above, we would like to know what is the
matrix $S$ when a constant gauge field $F$ is present on the
world-volume. For this purpose, we consider the corresponding
overlap conditions that the boundary state must satisfy
\cite{Callan:1986bc} \bea \left[\left(1+ \hat{F}
\right)^\alpha\,_\beta \,\alpha^\beta_n + \left( 1 -
\hat{F}\right)^\alpha\,_\beta \,{\tilde
\alpha}^\beta_{-n}\right]|B_X\rangle &=& 0 \nn \left(q^i -
y^i\right) |B_X\rangle = \left(\alpha^i_n - {\tilde
\alpha}^i_{-n}\right) |B_X\rangle &=& 0 \qquad n \neq 0\eea for the
bosonic part, and \bea \left[\left(1 + \hat{
F}\right)^\alpha\,_\beta \,\psi^\beta_m - {\rm i}\, \eta \left(1 -
\hat{F}\right)^\alpha\,_\beta \,{\tilde \psi}^\beta_{- m}\right]
|B_\psi, \eta\rangle &=& 0\nn \left(\psi_m^i +  {\rm i}\, \eta\,
{\tilde \psi}^i_{- m}\right) |B_\psi, \eta\rangle & = & 0 \eea for
the fermionic part. In the above, the Greek indices $\alpha, \beta,
\cdots$ label the world-volume directions $0, 1, \cdots, p$ along
which the D$_p$ brane extends, while the Latin indices $i, j,
\cdots$ label the directions transverse to the brane, i.e., $p + 1,
\cdots, 9$. We also define $\hat F = 2\pi \alpha' F$. One can check
that the above equations are solved by the ``coherent states"
(6)-(8) with the following matrix $S$
\cite{Callan:1986bc,DiVecchia:1999uf} \be S = \left(\left[(\eta -
\hat{F})(\eta + \hat{F})^{-1}\right]_{\alpha\beta},  -
\delta_{ij}\right) \ee and with the zero-mode parts given by \be
|B_X\rangle^{(0)} = \sqrt{- \det \left(\eta + \hat F\right)}
\,\delta^{9 - p} (q^i - y^i) \prod_{\mu = 0}^9 |k^\mu = 0\rangle\ee
for the bosonic sector, and by \be |B_\psi, \eta\rangle_{\rm
R}^{(0)} = \left(C \Gamma^0 \Gamma^1 \cdots \Gamma^p \frac{1 + {\rm
i}\, \eta \Gamma_{11}}{1 +  {\rm i}\, \eta } U \right)_{AB}
|A\rangle |\tilde B\rangle \ee for the R sector. In the above, we
have denoted by $y^i$ the positions of the D-brane along the
transverse directions, by $C$ the charge conjugation matrix and by
$U$ the following matrix \be U = \frac{1}{\sqrt{- \det (\eta + \hat
F)}} ; {\rm exp}\left(- \frac{1}{2} {\hat
F}_{\alpha\beta}\Gamma^\alpha\Gamma^\beta\right);\ee where the
symbol $;\quad ;$ means that one has to expand the exponential and
then to anti-symmetrize the indices of the $\Gamma$-matrices.
$|A\rangle |\tilde B\rangle$ stands for the spinor vacuum of the R-R
sector. We would like to point out that the $\eta$ in the above
means either sign $\pm$ or the flat signature matrix $(-1, +1,
\cdots, +1)$ on the world-volume and should not be confused from the
content.

One remark follows that the overlap equations (9) and (10) do not
allow to determine the overall normalization of the boundary state,
and not even to get the Born-Infeld prefactor of equation (12). The
latter was derived in \cite{Callan:1986bc}. It can also be obtained
by boosting the boundary state and then performing a T-duality as
explicitly shown in \cite{Billo:1997eg}. Notice also that this
prefactor is present only in the NS-NS component of the boundary
state because in the R-R sector it cancels out if we use the
explicit expressions for the matrix $U$ given in (14).

We also would like to point out that when we set the constant
world-volume field $F = 0$ in the above, everything will go over to
the case of a static D$_p$ brane without an external world-volume
field \cite{Di Vecchia:1997pr}. When the constant world-volume field
is an external electric field,  the corresponding boundary state
represents the BPS non-threshold (F, D$_p$) bound state where the
fundamental strings are represented by the electric flux. When the
external field is a magnetic one, the boundary state is then the BPS
non-threshold (D$_{p - 2}$, D$_p$) bound state where the lower
dimensional D$_{p - 2}$ branes are represented by the magnetic flux.
Each of the bound states preserves one half of the spacetime
supersymmetry of the underlying string theories. These two
non-threshold bound states are the focus of the present paper and we
will discuss their couplings to the massless modes of the type II
theories next.

\subsection{The couplings with bulk massless modes}

In this subsection, we will calculate the couplings of the
non-threshold (F, D$_p$) (or (D$_{p - 2}$, D$_p$)) bound state with
the bulk massless modes of the underlying type II theories through
the corresponding bound state world-volume effective action and bulk
effective action of the given string theory (IIA or IIB). We will
show that the couplings derived in the following agree completely
with those found through the boundary state approach given in
\cite{DiVecchia:1999uf}. By this, we also set the conventions for
the bulk fields in canonical form so the couplings can be used
correctly in finding the long-distance interaction between two
non-threshold bound states in the next section.

Let us first express the bulk fields in the effective action of a
given string theory in canonical form and for this purpose we need
only to consider the corresponding  bosonic action. Since this works
the same way in either IIA or IIB theory, we take IIA for
illustration. The bosonic part of the IIA low-energy effective
action in string frame is \bea S_{\rm IIA} &=& S_{\rm NS} + S_{\rm
R} + S_{\rm CS}, \\ S_{\rm NS} &=& \frac{1}{2 \kappa^2_{10}}\int
d^{10} x \sqrt{- G}\, e^{- 2 \Phi} \left[R + 4 (\nabla \Phi)^2 -
\frac{1}{2} |H_3|^2\right],\\ S_{\rm R} &=& - \frac{1}{4
\kappa^2_{10}}\int d^{10} x \sqrt{- G} \left[|F_2|^2 +
|\widetilde{F}_4|^2\right], \\ S_{\rm CS} &=& - \frac{1}{4
\kappa_{10}^2} \int B_2\wedge F_4 \wedge F_4,\eea  where NS-NS field
$H_3 = d B_2$ while the R-R fields $F_2 = d C_1, \widetilde{F}_4 = d
C_3 - C_1\wedge H_3$, respectively. In the above, we have grouped
terms according to whether the fields are in the NS-NS or R-R sector
except for the Chern-Simons action which contains both. The constant
$2\kappa_{10}^2$ appearing in the action is \be 2 \kappa_{10}^2 =
(2\pi)^7 \alpha'^4. \ee Since we are considering the field theory
limit, it is proper to express the above action in the Einstein or
canonical frame. This can be achieved through the so-called Einstein
metric $g_{\mu\nu}$ which is related to the string metric
$G_{\mu\nu}$ through the following
 \be g_{\mu\nu} = e^{- \phi/2} G_{\mu\nu}\ee where we have
defined \be \phi = \Phi - \Phi_0 \ee with $\Phi_0$  the asymptotic
value (or VEV ) of the dilaton. In this frame, we have \bea S_{\rm
NS} &=& \frac{1}{2 g_s^2 \kappa^2_{10}}\int d^{10} x \sqrt{- g}
\left[R - \frac{1}{2} (\nabla \phi)^2 - \frac{1}{2} e^{- \phi}
|H_3|^2\right],\\ S_{\rm R} &=& - \frac{1}{4 \kappa^2_{10}}\int
d^{10} x \sqrt{- g} \left[e^{ 3 \phi/2} |F_2|^2 + e^{\phi/2}
|\widetilde{F}_4|^2\right], \eea while the $S_{\rm CS}$ remains the
same.  In the above, we have introduced the string coupling $g_s =
e^{\Phi_0}$ and  with this the physical gravitational coupling is
\be 2 \kappa^2 = 2 g_s^2 \kappa_{10}^2.\ee Considering small
fluctuations of fields with respect to the flat Minkowski
background, we have the action \be S_{\rm IIA} = \frac{1}{2\kappa^2}
\int d^{10} x \left[- \frac{1}{4} \nabla\, h^{\mu\nu} \nabla \,
h_{\mu\nu} - \frac{1}{2} (\nabla\phi)^2 - \frac{1}{2} |H_3|^2\right]
- \frac{1}{4 \kappa_{10}^2} \int d^{10} x \left[ |F_2|^2 + |F_4|^2
\right] + \cdots \ee where we keep only the lowest order terms and
$\cdots$ represents the higher order terms. In the above, $F_2$ and
$H_3$ have their respective definitions defined earlier,  $F_4 = d
A_3$, and we have expanded \be g_{\mu\nu} = \eta_{\mu\nu} +
h_{\mu\nu}\ee with $\eta_{\mu\nu}$ the flat metric and the usual
harmonic gauge for $h_{\mu\nu}$. The above action obviously becomes
canonical with the following scalings: \be h_{\mu\nu} \rightarrow 2
\kappa \,h_{\mu\nu}, \quad \phi \rightarrow \sqrt{2} \kappa\,
\phi,\quad B_{\mu\nu} \to \sqrt{2} \kappa \,B_{\mu\nu}\ee for NS-NS
fields and \be C_{n} \rightarrow \sqrt{2} \kappa_{10}\, C_{n}\ee for
rank-n R-R potential. Note that the scaling of a NS-NS field differs
from that of a R-R potential by a string coupling $g_s$ except for
the graviton which has an additional factor of $\sqrt{2}$. These
will help us to determine the corresponding couplings of bulk fields
with the D-brane in the canonical form to which we will turn  next.

For this, let us consider the bosonic world-volume action of a D$_p$
brane with a constant world-volume field $\hat F$ in string frame
which is \be S = - T_p \int d^{1 + p} \sigma \,e^{- \Phi}\, \sqrt{-
\det(G + B + \hat F)} + T_p \int \left[e^{B + \hat F} \wedge \sum_l
C_{p + 1 - 2 l}\right]_{p + 1},\ee where the metric $G$, the NS-NS
rank-2 potential $B$ and the R-R potential $C_{p + 1 - 2l}$ are the
pullbacks of the corresponding bulk fields to the world-volume, a
n-form potential is defined as \be A_{n} = \frac{1}{n!} A_{\alpha_1
\cdots \alpha_n} d\sigma^{\alpha_1}\wedge \cdots \wedge
d\sigma^{\alpha_n},\ee and \be T_p = \frac{2\pi}{(4\pi^2
\alpha')^{\frac{p + 1}{2}}}. \ee The square bracket in the above
Wess-Zumino term means that in expanding the exponential form one
picks up only terms of total degree of $(p + 1)$.   We now express
the above action in Einstein frame using equation (20) as \be S = -
\frac{T_p}{g_s} \int d^{1 + p} \sigma \,e^{\frac{ (p - 3)\phi}{4}}
\sqrt{- \det[g + (B + \hat F) e^{- \phi/2}]} + T_p \int \left[e^{B +
\hat F} \wedge \sum_l C_{p + 1 - 2 l}\right]_{p + 1},\ee where we
have also used eq.(21). By the same token, we expand the above
action with fixed $\hat F$ to the leading order in small
fluctuations of background as before and we end up with \bea S =& -&
\frac{T_p}{g_s} \int d^{1 + p} \sigma\, \sqrt{- \det(\eta + \hat
F)}\left\{1 + \frac{1}{2} [(\eta + \hat F)^{-1}]^{\alpha\beta} (
h_{\beta\alpha} + B_{\beta\alpha}) \right. \nn  &+& \left.
\frac{1}{4}[ p - 3 - {\rm Tr}(\hat F (\eta + \hat F)^{-1}) ] \phi
\right\} + T_p \int \left(C_{p + 1} + \hat F \wedge C_{p - 1} +
\cdots \right), \eea where $\cdots$ means terms with the lower rank
of R-R potentials wedged with more $\hat F$'s. Now we use the
scalings in (27) for NS-NS fields and in (28) for R-R fields to
replace the background fluctuations in the above action and have
\bea S =& -& \frac{T_p \kappa}{g_s} \int d^{1 + p} \sigma\, \sqrt{-
\det(\eta + \hat F)}\left\{1 + \frac{1}{\sqrt{2}} [(\eta + \hat
F)^{-1}]^{\alpha\beta} ( \sqrt{2}\, h_{\beta\alpha} +
B_{\beta\alpha}) \right. \nn  &+& \left. \frac{1}{2\sqrt{2}}[ p - 3
- {\rm Tr}(\hat F (\eta + \hat F)^{-1}) ] \phi \right\} + \sqrt{2}
T_p \kappa_{10} \int \left(C_{p + 1} + \hat F \wedge C_{p - 1} +
\cdots \right). \eea From the above action we can read the
respective coupling in the canonical form  \be J_h = - c_p V_{p + 1}
\sqrt{- \det(\eta + \hat F)} \left[(\eta + \hat F)^{-1}
\right]^{\alpha\beta} h_{\beta\alpha} \ee  for the graviton,  \be
J_\phi = \frac{c_p}{2\sqrt{2}} V_{p + 1}\sqrt{- \det(\eta + \hat F)}
\left[3 - p + {\rm Tr}\left(\hat F (\eta + \hat
F)^{-1}\right)\right]\,\phi\ee for the dilaton, \be J_B = -
\frac{c_p}{\sqrt{2}} V_{p + 1} \sqrt{- \det(\eta + \hat F)}
\left[(\eta + \hat F)^{-1} \right]^{\alpha\beta} B_{\beta\alpha}\ee
for the Kalb-Ramond field,  \be J_{C_{p + 1}} = \frac{\sqrt{2}\,
c_p}{(p + 1)!} V_{p + 1} C_{\alpha_0\alpha_1\cdots\alpha_p}
\varepsilon^{\alpha_0\alpha_1\cdots\alpha_p}\ee for the  $(p +
1)$-from RR potential,  \be J_{C_{p - 1}} = \frac{\sqrt{2}\, c_p}{2
(p - 1)!} V_{p + 1} {\hat F}_{\alpha_0\alpha_1} C_{\alpha_2 \cdots
\alpha_p}\varepsilon^{\alpha_0\alpha_1\cdots\alpha_p}\ee for the $(p
- 1)$-form R-R potential and so on. In the above, $V_{p + 1}$ is the
world-volume of the brane, $\varepsilon^{\alpha_0 \cdots \alpha_p}$
is the totally antisymmetric tensor on the D-brane
world-volume\footnote{By conventions, $\varepsilon^{0,1,\cdots, p} =
- \varepsilon_{0,1,\cdots, p} = 1$.} , and we assume that the
background field fluctuations depend on only the transverse
coordinates to the static brane. In the above, we have used $c_p =
T_p \kappa/g_s = T_p \kappa_{10}$ with the aid of Eqs.(5), (19),
(24) and (31). The couplings obtained above are in complete
agreement with those obtained in \cite{DiVecchia:1999uf} using
boundary state approach and will be used in the following section to
obtain the long-range forces between two non-threshold bound states.

\section{The long-range interactions}

In this section, we will calculate the lowest-order contribution
to the interaction between two arbitrary (F, D$_p$) ( or (D$_{p -
2}$, D$_p$)) bound states placed parallel to each other at a given
separation $Y$ due to the exchanges of massless modes, therefore
representing the force at large separation. As mentioned in the
Introduction, the lower dimensional brane in the bound state can
be represented by the corresponding flux on the D$_p$ brane
world-volume. For the present case, the F-strings in (F, D$_p$)
can be represented by an electric flux along a given direction on
the p-brane worldvolune while the D$_{p - 2}$ branes in (D$_{p -
2}$, D$_p$) can be represented by a magnetic flux similarly.

Let us begin with the non-threshold (F, D$_p$) states. We choose
the constant electric flux $\hat F$ the following way \be \hat F =
\left(\begin{array}{cccccc}
0& -f& & &&\\
f& 0&  & && \\
 &  &\cdot&&& \\
 &  &     &\cdot&&\\
 &  &     &     &\cdot& \\
 &  &     &     &     &0 \end{array} \right)_{(p + 1) \times (p +
 1)}.\ee
The couplings derived in the previous section are for a single
 D$_p$ brane in the bound state and for multiple D$_p$ branes, we
 should replace the $c_p$ by n $c_p$ in the couplings with $n$ an integer. The
 constant flux is also quantized and is given for an electric flux as
 \cite{DiVecchia:1999uf} \be - \frac{n f}{\sqrt{1 - f^2}} = m \, g_s \ee with m an
 integer. This gives $f = - m /\triangle^{1/2}_{(m,n)}$
 where we have defined \be \triangle_{(m, n)} = m^2 +
 \frac{n^2}{g_s^2}.\ee
 Then we have \be - \det (\eta + \hat F) = 1 - f^2 =
 \frac{n^2/g_s^2}{\triangle_{(m,n)}}\ee and
 \bea V \equiv (\eta + \hat F)^{-1} &=& \left(\begin{array}{ccccccc}
- \frac{1}{1 - f^2} & - \frac{f}{1 - f^2} & & &&&\\
\frac{f}{1 - f^2}& \frac{1}{1 - f^2} &  & && &\\
 &   & 1&&&&\\
 &  &    &\cdot&&& \\
 &  &     & &\cdot&&\\
 &  &     &     &&\cdot& \\
 &  &     &     &     &&1 \end{array} \right)_{(p + 1) \times (p +
 1)} \nn
 &=& \left(\begin{array}{ccccccc}
- \frac{g_s^2 \triangle_{(m,n)}}{n^2}  &  \frac{m g^2_s \triangle^{1/2}_{(m,n)} }{n^2} && & &&\\
- \frac{m g^2_s \triangle^{1/2}_{(m,n)}}{n^2}& \frac{g_s^2 \triangle_{(m,n)}}{n^2} & & & && \\
 &  & 1 &&&&\\
 &  &    &\cdot&&& \\
 &  &    &      &\cdot&&\\
 &  &     &     &      &\cdot& \\
 &  &     &     &     & & 1 \end{array} \right)_{(p + 1) \times (p +
 1)}.\eea
With the above, we have now the couplings using (35)-(39) as \bea
J^i_h &=& - c_p V_{p + 1} \, \frac{n_i^2}{g_s \triangle^{1/2}_{(m_i,
n_i)}} \, V_i^{\alpha\beta} h_{\beta\alpha},\quad
 J^i_\phi =
 \frac{c_p}{2\sqrt{2}} V_{p + 1} \, \frac{(3 - p) n_i^2 - 2 m_i^2 g^2_s }{g_s \triangle^{1/2}_{(m_i,
n_i)}}\,\phi \nn
 J^i_B &=& - \frac{c_p}{\sqrt{2}} V_{p + 1}\,\frac{n_i^2 }{g_s \triangle^{1/2}_{(m_i, n_i)}}\, V_i^{\alpha\beta}
B_{\beta\alpha},\eea for the NS-NS fields and \be J^i_{C_{p + 1}} =
\sqrt{2}\, c_p\, V_{p + 1} \, n_i \, C_{0 1\cdots p}, \quad
J^i_{C_{p - 1}} = c_p\, V_{p + 1}\,\frac{\sqrt{2} \,n_i\,
m_i}{\triangle^{1/2}_{(m_i, n_i)}} \, C_{2 3 \cdots p} \ee for the
R-R fields. Here $i$ denotes the respective bound state with $i =1,
2$.

We now calculate the long-range interaction (in momentum space)
between two parallel (F, D$_p$) bound states separated by a
transverse distance $Y$ with each state characterized by a pair of
integers $(m_i, n_i)$, respectively. The gravitational contribution
due to the exchange of
 graviton is \be U_h = \frac{1}{V_{p + 1}}
\underbrace{J^{(1)}_h J^{(2)}_h} = c_p^2 V_{p + 1} \frac{n_1^2
n_2^2}{g_s^2 \triangle^{1/2}_{(m_1, n_1)} \triangle^{1/2}_{(m_2,
n_2)}} V_1^{\alpha\beta} V_2^{\gamma\delta}
\underbrace{h_{\beta\alpha} h_{\delta\gamma}}\ee where the
propagator reads \be \underbrace{h_{\beta\alpha} h_{\delta\gamma}} =
\left[\frac{1}{2}\left(\eta_{\beta\delta} \eta_{\alpha\gamma} +
\eta_{\alpha\delta} \eta_{\beta\gamma}\right) - \frac{1}{8}
\eta_{\alpha\beta} \eta_{\gamma\delta}\right] \frac{1}{k_\bot^2}\ee
for the canonically normalized graviton propagating in the
transverse directions in the de Donder (harmonic) gauge. The
explicit expression for the interaction can be obtained using the
matrix $V$ in the second line of (44) as \be U_h = \frac{c_p^2}{8
g^2_s} \frac{ V_{p + 1}}{k^2_\bot} \frac{12 g_s^4 m_1^2 m_2^2 + 2 (7
- p) g_s^2 (m_1^2 n_2^2 + m_2^2 n_1^2) + (7 - p) (p + 1) n_1^2
n_2^2} {\Omega} \ee with \be \Omega \equiv \triangle^{1/2}_{(m_1,
n_1)} \triangle^{1/2}_{(m_2, n_2)} = \sqrt{(m_1^2 +
\frac{n_1^2}{g_s^2})(m_2^2 + \frac{n^2_2}{g_s^2})}.\ee The
contribution to the interaction due to the exchange of dilaton can
be calculated as \be U_\phi =\frac{1}{V_{p + 1}}
\underbrace{J^1_\phi J^2_\phi} = \frac{c_p^2}{8 g_s^2} V_{p + 1}
\frac{4 g^4_s m_1^2 m_2^2 - 2 (3 - p) g_s^2 (m_1^2 n_2^2 + n_1^2
m_2^2) + (3 - p)^2 n_1^2 n_2^2}{\Omega}\, \underbrace{\phi \phi},
\ee where $\Omega$ is given in eq.(50) and the dilaton propagator is
\be \underbrace{\phi \phi } = \frac{1}{k^2_\bot}.\ee So we have \be
U_\phi = \frac{c_p^2}{8 g_s^2} \frac{V_{p + 1}}{k^2_\bot} \frac{4
m_1^2 m_2^2 - 2 (3 - p) g_s^2 (m_1^2 n_2^2 + n_1^2 m_2^2) + (3 -
p)^2 n_1^2 n_2^2}{\Omega}.\ee The contribution due to the exchange
of Kalb-Ramond field can be calculated similarly as \be U_B =
\frac{1}{V_{p + 1}} \underbrace{J^1_B J^2_B} = \frac{c^2_p}{2 g_s^2}
V_{p + 1} \frac{n_1^2 n_2^2}{\Omega} V_1^{\alpha\beta}
V_2^{\gamma\delta} \underbrace{B_{\beta\alpha} B_{\delta\gamma}}.
\ee Using the propagator for the Kalb-Ramond field  \be
\underbrace{B_{\beta\alpha} B_{\delta\gamma}} =
\left(\eta_{\beta\delta} \eta_{\alpha\gamma} -
\eta_{\alpha\delta}\eta_{\beta\gamma}\right)\frac{1}{k^2_\bot}\ee
and the explicit expression for the matrices $V_i$, we have \be U_B
= \frac{c^2_p}{8 g_s^2} \frac{V_{p + 1}}{k^2_\bot} (- 16 m_1 m_2
g_s^4).\ee

We now turn to the calculations of the contributions from R-R
fields. The contribution from the exchange of R-R potential
$C_{01\cdots p}$ is \be U_{C_{p + 1}} \equiv \frac{1}{V_{p + 1}}
\underbrace{J^1_{C_{p + 1}} J^2_{C_{p + 1}}} = 2 c_p^2 V_{p + 1} n_1
n_2 \underbrace{C_{01\cdots p} C_{01\cdots p}}.\ee Using the
propagator for the rank-(p + 1) R-R potential \be
\underbrace{C_{01\cdots p} C_{01\cdots p}} = -
\frac{1}{k^2_\bot},\ee we have \be U_{C_{p + 1}} = \frac{c_p^2}{8
g_s^2} \frac{V_{p + 1}}{k^2_\bot} ( - 16 n_1 n_2 g^2_s).\ee
Similarly we have \bea U_{C_{p - 1}} &\equiv& \frac{1}{V_{p + 1}}
\underbrace{J^1_{C_{p - 1}} J^2_{C_{p - 1}}} = 2 c^2_p V_{p + 1}
\frac{m_1 m_2 n_1 n_2}{\Omega} \underbrace{C_{23\cdots p}
C_{23\cdots p}}\nn &=& \frac{c_p^2}{8 g_s^2} \frac{V_{p +
1}}{k^2_\bot} \frac{16 m_1 m_2 n_1 n_2 g_s^2}{\Omega},\eea where we
have used the propagator for the rank-(p - 1) R-R potential \be
\underbrace{C_{23\cdots p} C_{23\cdots p}} = \frac{1}{k^2_\bot}.\ee
Note that apart from the overall factor $c_p^2 \frac{V_{p +
1}}{k^2_\bot}$, the form field contributions are independent of the
dimensionality of the bound state while this is not case for either
the graviton or the dilaton contribution.

We would like to point out that each of the components calculated
above agrees completely with what has been given in
\cite{DiVecchia:1999uf} when we set $(m_1, n_1) = (m_2, n_2)$ and
$g_s = 1$, i.e., when the two bound states are identical with string
coupling set to one. We here generalize the calculations there for
two arbitrary bound states which are characterized by their
respective pair of integers $(m_i, n_i)$ with $i = 1, 2$. The total
contribution to the interaction from the NS-NS sector is \bea U_{\rm
NS-NS} &=& U_h + U_\phi + U_B\nn &=& c_p^2\, \frac{V_{p +
1}}{k^2_\bot}\frac{2 g_s^4 m_1^2 m_2^2 + g_s^2 (m_1^2 n_2^2 + m_2^2
n_1^2) + 2 n_1^2 n_2^2 - 2 m_1 m_2 g^4_s \Omega}{g_s^2 \Omega}\nn
&=& c_p^2\,\frac{V_{p + 1}}{k^2_\bot} U_{\rm NS} (m_1, n_1; m_2,
n_2) \eea where in the last line we have made use of the explicit
expression for $\Omega$ given in eq.(50) and \be U_{\rm NS} (m_1,
n_1; m_2, n_2) = \frac{g_s^4 m_1^2 m_2^2 + n_1^2 n_2^2 + g_s^4
\Omega^2 - 2 m_1 m_2 g_s^4 \Omega}{g^2_s \Omega}, \ee while the
total R-R contribution is \be U_{\rm R-R} = U_{C_{p + 1}} + U_{C_{p
- 1}} = - c_p^2\, \frac{V_{p + 1}}{k^2_\bot} U_{\rm R} (m_1, n_1;
m_2, n_2) \ee where \be U_{\rm R} (m_1, n_1; m_2, n_2) = \frac{2 n_1
n_2 (\Omega - m_1 m_2) g_s^2}{g_s^2 \Omega}. \ee Note that although
either the graviton or the dilaton contribution apart from the
factor $c_p^2 \frac{V_{p + 1}}{k^2_\bot}$ depends on the
dimensionality of the brane, their addition is not. This has to be
so since the form field contributions are independent of the
dimensionality and ``no-force" condition holds once we set the two
bound states identical. The total contribution from both sectors is
\bea U &=& U_{\rm NS-NS} + U_{\rm R-R}\nn & =& c_p^2\, \frac{V_{p +
1}}{k^2_\bot}\frac{\left[(g_s^2 m_1 m_2 + n_1 n_2) - g_s^2
\Omega\right]^2 }{g_s^2 \Omega} \ge 0. \eea This clearly shows that
the interaction is in general attractive\footnote{We choose
conventions here that $U > 0$ means attractive which differs from
standard one by a sign.} and vanishes only if \be g_s^2 m_1 m_2 +
n_1 n_2 = g_s^2 \Omega > 0.\ee For non-degenerate case, i.e., $m_i
n_i \neq 0$ with $i = 1, 2$,  the above implies $m_1/n_1 = m_2/n_2$
and $n_1 n_2 > 0$. In showing this, we have made use of the explicit
expression for $\Omega$ given in (50). The vanishing result for the
special case of $(m_1, n_1) = (m_2, n_2)$ was previously shown in
\cite{DiVecchia:1999uf} and we here  generalize it to a general
case.

We now turn to the case for the non-threshold (D$_{p - 2}$, D$_p$)
bound state. The calculations are similar and we list below only
the necessary steps and the main results. The constant magnetic
flux $\hat F$ on the world-volume is chosen as \be \hat F =
\left(\begin{array}{cccccc}
0& & & &&\\
& \cdot&  & && \\
 &  &\cdot&&& \\
 &  &     &\cdot&&\\
 &  &     &     &0& -f\\
 &  &     &     &  f   &0 \end{array} \right)_{(p + 1) \times (p +
 1)}.\ee
Here again we need to replace the $c_p$ for a single D$_p$ brane
in the bound state by $n c_p$ for multiple branes with $n$ an
integer (also due to charge quantization) in the couplings. The
constant magnetic flux is also quantized and in the present case
is given by $ - n f = m$ which gives $f = - m/n$. So we have now
\be - \det (\eta + \hat F) = 1 + f^2 = \frac{n^2 + m^2}{n^2}. \ee
and \bea V \equiv  (\eta + \hat F)^{ - 1} &=&
\left(\begin{array}{ccccccc}
- 1 &  & & &&&\\
& 1 &  & && &\\
 &   & \cdot&&&&\\
 &  &    &\cdot&&& \\
 &  &     & &\cdot&&\\
 &  &     &     &&\frac{1}{1 + f^2}&\frac{f}{1 + f^2} \\
 &  &     &     &     &- \frac{f}{1 + f^2}&\frac{1}{1 + f^2} \end{array} \right)_{(p + 1) \times (p +
 1)} \nn
 &=& \left(\begin{array}{ccccccc}
- 1 &  & & &&&\\
& 1 &  & && &\\
 &   & \cdot&&&&\\
 &  &    &\cdot&&& \\
 &  &     & &\cdot&&\\
 &  &     &     &&\frac{n^2}{m^2 + n^2}& - \frac{n m }{m^2 + n^2} \\
 &  &     &     &     & \frac{n m}{n^2 + m^2}&\frac{n^2}{n^2 + m^2} \end{array} \right)_{(p + 1) \times (p +
 1)}.\eea
We then have the explicit couplings for the respective bound state
denoted by index $i$ with $i = 1, 2$ from (35)-(39) in the previous
section as \bea J^i_h &=& - c_p V_{p + 1} \sqrt{m_i^2 + n_i^2}\,
V_i^{\alpha\beta} h_{\beta\alpha},\quad J^i_\phi = \frac{c_p}{2\sqrt
2} V_{p + 1} \frac{(3 - p)(n_i^2 + m_i^2) + 2 m_i^2}{\sqrt{m_i^2 +
n_i^2}}\, \phi\nn J^i_B & = & - \frac{c_p}{\sqrt{2}} V_{p + 1}
\sqrt{m_i^2 + n_i^2} \,V_i^{\alpha\beta} B_{\beta\alpha} \eea for
the NS-NS couplings and \be J^i_{C_{p + 1}} = \sqrt{2}\,c_p\, V_{p +
1} n_i\, C_{01\cdots p},\quad J^i_{C_{p - 1}} =  \sqrt{2}\, c_p\,
V_{p + 1} m_i \,C_{01\cdots p - 2}\ee for the R-R couplings. We then
have the long-range interaction due to the exchange of each of the
massless fields respectively as \bea U_\phi &=& \frac{c_p^2}{8}
\frac{V_{p + 1}}{k^2_\bot} \frac{(5 - p)^2 m_1^2 m_2^2 + (5 - p)(3 -
p) (m_1^2 n_2^2 + n_1^2 m_2^2) + (3 - p)^2 n_1^2
n_2^2}{\tilde\Omega},\nn
 U_h &=& \frac{c_p^2}{8} \frac{V_{p +
1}}{k^2_\bot} \frac{(9 - p)(p - 1) m_1^2 m_2^2 + (7 - p)(p -
1)(m_1^2 n_2^2 + n_1^2 m_2^2) + (7 - p)(p + 1) n_1^2 n_2^2}{\tilde
\Omega},\nn U_B &=& \frac{c_p^2}{8} \frac{V_{p + 1}}{k^2_\bot}
\frac{16 m_1 m_2 n_1 n_2 }{\tilde \Omega} \eea for the NS-NS fields
and \be U_{C_{p + 1}} = c_p^2\, \frac{V_{p + 1}}{k^2_\bot} (- 2 n_1
n_2),\quad U_{C_{p - 1}} = c_p^2\, \frac{V_{p + 1}}{k^2_\bot} (- 2
m_1 m_2)\ee for the R-R fields. In the above, we have defined \be
\tilde \Omega = \sqrt{(m_1^2 + n_1^2)(m_2^2 + n_2^2)}.\ee We again
have that the interaction contribution due to the exchange of the
dilaton or the graviton in the NS-NS sector apart from the facotr
$c_p^2 \frac{V_{p + 1}}{k^2_\bot}$ still depends on the
dimensionality of the world-volume while this is not the case for
any form field in either NSNS secotr or the R-R sector. The total
contribution to the interaction from the NS-NS sector is \bea U_{\rm
NS-NS} &=& U_\phi + U_h + U_B \nn &=& c_p^2\, \frac{V_{p +
1}}{k^2_\bot} U_{\rm NS} (m_1, n_1; m_2, n_2),\eea where \be U_{\rm
NS} (m_1, n_1; m_2, n_2) = \frac{2 m_1^2 m_2^2 + (m_1^2 n_2^2 +
n_1^2 m_2^2) + 2 n_1^2 n_2^2 + 2 m_1 m_2 n_1 n_2}{\tilde \Omega},
\ee independent of the dimensionality of the world-volume. The total
interaction from the R-R sector is \be U_{\rm R-R} = U_{C_{p + 1}} +
U_{C_{p - 1}} = - c_p^2\, \frac{V_{p + 1}}{k^2_\bot} U_{\rm R} (m_1,
n_1; m_2, n_2), \ee where  \be U_{\rm R} (m_1, n_1; m_2, n_2) = 2
(n_1 n_2 + m_1 m_2).\ee The total interaction from both sectors is
now \bea U &=& U_{\rm NS-NS} + U_{\rm R-R}\nn &=& c_p^2\, \frac{V_{p
+ 1}}{k^2_\bot} \frac{(m_1 m_2 + n_1 n_2 - \tilde \Omega)^2}{\tilde
\Omega} \geq 0 \eea where in the second line we have used the
explicit expression for $\tilde \Omega$ given in eq.(75). This also
clearly shows that the interaction is in general attractive and
vanishes only if \be m_1 m_2 +  n_1 n_2 = \tilde \Omega\ee which
again implies $m_1/n_1 = m_2/n_2$ and $n_1 n_2 > 0$ for the
non-degenerate case, i.e., $m_i n_i \neq  0$ with $i = 1, 2$, the
expected supersymmetry preserving result.

We can use Fourier transformation to obtain the corresponding
interaction in coordinate space when $p < 7$ as \be U (Y) = \int
\frac{d^\bot k_\bot}{ (2\pi)^\bot} e^{- {\rm i} k_\bot \cdot Y} U
(k_\bot) = \frac{C (m_1, n_1; m_2, n_2)} {Y^{7 - p}}\ee where \be C
(m_1, n_1; m_2, n_2) = \frac{c_p^2\, V_{p + 1} \,U (m_1, n_1; m_2,
n_2) }{(7 - p) \Omega_{8 - p}}\ee with \be U (m_1, n_1; m_2, n_2)  =
\left\{\begin{array} {cc} \frac{[(g_s^2 m_1 m_2 + n_1 n_2) - g_s^2
\Omega]^2}{g_s^2 \Omega} & {\rm for \, the \, case\, of}
\,(F, D_p), \\
&\\
 \frac{[(m_1 m_2 + n_1 n_2) - \tilde \Omega]^2}{\tilde \Omega} &
{\rm for\, the \, case \, of}\, (D_{p -2}, D_p),
\end{array}\right. \ee and $Y^2 = Y_i Y^i$ with the summation index
 $i$ along the transverse
directions. In the above, we have used the following relation \be
\int \frac{d^\bot k_\bot}{(2\pi)^\bot} \frac{e^{- {\rm i} k_\bot
\cdot Y}}{k^2_\bot} = \frac{1}{(7 - p) Y^{7 - p} \Omega_{8 - p}},\ee
where $\Omega_q = 2 \pi^{(q + 1)/2}/\Gamma((q + 1)/2)$ is the volume
of unit q-sphere.

\section{The string-level force calculations}

We want to go one step further to calculate the forces between two
(F, D$_p$) or (D$_{p - 2}$, D$_p$) bound states at a separation $Y$
at the string level as the corresponding interaction vacuum
amplitude\footnote{Actually it is the vacuum free energy.}. In
addition, we will use the results to discuss certain properties of
the underlying systems such as the analytic structure of the
amplitude and to calculate the rate of pair production of open
strings in the open string channel.

The interaction vacuum amplitude can be calculated via
 \be \Gamma
= \langle B (m_1, n_1) | D |B (m_2, n_2) \rangle \ee where the bound
state with a constant
 world-volume field in each sector has been given in section 2 and
 is characterized by a pair of integers $(m_i, n_i)$ with $i = 1,
 2$
and $D$ is the closed string propagator defined as \be D =
\frac{\alpha'}{4 \pi} \int_{|z|^2 \le 1} \frac{d^2 z}{|z|^2} z^{L_0}
 {\bar z}^{{\tilde L}_0}.\ee Here $L_0$ and ${\tilde L}_0$ are
the respective left and right mover total zero-mode Virasoro
generators of matter fields, ghosts and superghosts. For example,
$L_0 = L^X_0 + L_0^\psi + L_0^{gh} + L_0^{sgh}$ where $L_0^X,
L_0^\psi, L_0^{gh}$ and $L_0^{sgh}$ represent contributions from
matter fields $X^\mu$, matter fields $\psi^\mu$, ghosts $b$ and $c$,
and superghosts $\beta$ and $\gamma$, respectively, and their
explicit expressions can be found in any standard discussion of
superstring theories, for example in \cite{Di Vecchia:1999rh},
therefore will not be presented here even though we will need them
in our following calculations. The above total vacuum amplitude has
contributions from both NS-NS and R-R sectors, respectively, and can
be written as $\Gamma = \Gamma_{\rm NS} + \Gamma_{\rm R}$. In
calculating either $\Gamma_{\rm NS}$ or $\Gamma_{\rm R}$, we need to
keep in mind that the boundary state used should be the GSO
projected one as given in Eq. (1) or Eq. (2). For this purpose, we
need to calculate first the following amplitude \be \Gamma (\eta',
\eta) = \langle B^1, \eta'| D |B^2, \eta\rangle \ee in each sector
with $\eta' \eta = \pm$ and $B^i = B (m_i, n_i)$. In doing the
calculations, we can set  $\tilde L_0 = L_0$ in the above propagator
due to the fact that $\tilde L_0 |B\rangle = L_0 |B\rangle$, which
can be used to simplify the calculations. Given the structure of the
boundary state in Eq. (3) and Eq. (4), the amplitude $\Gamma (\eta',
\eta)$ can be factorized as \be \Gamma (\eta', \eta) = \frac{n_1 n_2
c_p^2}{4} \frac{\alpha'}{4 \pi} \int_{|z| \le 1} \frac{d^2 z}{|z|^2}
A^X \, A^{bc}\, A^\psi (\eta', \eta)\, A^{\beta\gamma} (\eta',
\eta),\ee where we have replaced the $c_p$ in the boundary state
given in section 2 by $n c_p$ with $n$ an integer to count the
multiplicity of the D$_p$ branes in the bound state. In the above,
\bea && A^X = \langle B^1_X | |z|^{2 L^X_0} |B^2_X \rangle,\qquad
A^\psi (\eta', \eta) = \langle B^1_\psi, \eta'| |z|^{2 L_0^\psi}
|B^2_\psi, \eta \rangle,\nn && A^{bc} = \langle B^1_{gh} | |z|^{2
L_0^{gh}} | B^2_{gh}\rangle, \qquad A^{\beta\gamma} (\eta', \eta) =
\langle B^1_{sgh}, \eta'| |z|^{2 L_0^{sgh}} |B^2_{sgh}, \eta\rangle.
\eea In order to perform the calculations using the boundary states
given in (6)-(8), (12) and (13), we need to specify the worldvolume
gauge field and the S-matrix given in (11) for both (F, D$_p$) and
(D$_{p - 2}$, D$_p$) bound states, respectively. For the case of (F,
D$_p$), we need to use (40) for $\hat F$ with $f$ determined by
(41), i.e., $f = - m /\triangle^{1/2}_{(m, n)}$ through the charge
quantization. The corresponding longitudinal part of the S matrix as
given in (11), is now \bea S_{\alpha\beta} &=&
\left(\begin{array}{ccccccc}
- \frac{1 + f^2}{1 - f^2} &  \frac{2f}{1 - f^2} & & &&&\\
- \frac{2 f}{1 - f^2}& \frac{1 + f^2}{1 - f^2} &  & && &\\
 &   & 1&&&&\\
 &  &    &\cdot&&& \\
 &  &     & &\cdot&&\\
 &  &     &     &&\cdot& \\
 &  &     &     &     &&1 \end{array} \right)_{(p + 1) \times (p +
 1)} \nn
 &=& \left(\begin{array}{ccccccc}
- \frac{g_s^2 (\triangle_{(m,n)} + m^2)}{n^2}  &  - \frac{2 m g^2_s \triangle^{1/2}_{(m,n)} }{n^2} && & &&\\
 \frac{2 m g^2_s \triangle^{1/2}_{(m,n)}}{n^2}& \frac{g_s^2 (\triangle_{(m,n)} + m^2)}{n^2} & & & && \\
 &  & 1 &&&&\\
 &  &    &\cdot&&& \\
 &  &    &      &\cdot&&\\
 &  &     &     &      &\cdot& \\
 &  &     &     &     & & 1 \end{array} \right)_{(p + 1) \times (p +
 1)}.\eea
While for (D$_{p - 2}$, D$_p$), we need to use (68) for $\hat F$
with the quantized $f = - m/n$. Now we have the longitudinal part of
the S matrix as \bea S_{\alpha\beta} &=&
\left(\begin{array}{ccccccc}
- 1 &  & & &&&\\
& 1 &  & && &\\
 &   & \cdot&&&&\\
 &  &    &\cdot&&& \\
 &  &     & &\cdot&&\\
 &  &     &     &&\frac{1 - f^2}{1 + f^2}&\frac{2 f}{1 + f^2} \\
 &  &     &     &     &- \frac{2 f}{1 + f^2}&\frac{1 - f^2}{1 + f^2} \end{array} \right)_{(p + 1) \times (p +
 1)} \nn
 &=& \left(\begin{array}{ccccccc}
- 1 &  & & &&&\\
& 1 &  & && &\\
 &   & \cdot&&&&\\
 &  &    &\cdot&&& \\
 &  &     & &\cdot&&\\
 &  &     &     &&\frac{n^2 - m^2}{m^2 + n^2}& - \frac{2 n m }{m^2 + n^2} \\
 &  &     &     &     & \frac{2 n m}{n^2 + m^2}&\frac{n^2 - m^2}{n^2 + m^2} \end{array} \right)_{(p + 1) \times (p +
 1)}.\eea
With the above preparations, we are now ready to perform rather
straightforward calculations for the various matrix elements
specified in (90) in either NS-NS or R-R sector for either of the
bound states under consideration, using (6)-(8), (12) and (13) for
the boundary states with $\hat F$ and the matrix $S$ given in (11)
as just described for either of the bound states. We have now \bea
&&A^X = C_F V_{p + 1} e^{- \frac{Y^2}{2\pi \alpha' t}} \left(2\pi^2
\alpha'\, t\right)^{- \frac{9 - p}{2}} \, \prod_{n = 1}^\infty
\frac{1}{(1 - \lambda |z|^{2n})(1 - \lambda^{-1} |z|^{2n})(1 -
|z|^{2n})^8},\nn && A^{bc} = |z|^{-2} \prod_{n = 1}^\infty (1 -
|z|^{2n})^2,\eea for both NS-NS and R-R sectors, \bea && A_{\rm
NS}^{\beta\gamma} (\eta', \eta) = |z| \prod_{n = 1}^{\infty}
\frac{1}{(1 + \eta' \eta\, |z|^{2n - 1})^2},\nn && A^\psi_{\rm NS} =
\prod_{n = 1}^\infty (1 + \eta' \eta \,\lambda |z|^{2n - 1}) (1 +
\eta' \eta\,\lambda^{-1}\, |z|^{2n - 1}) (1 + \eta' \eta \,|z|^{2n -
1})^8, \eea for NS-NS sector, and \be A^{\beta\gamma}_{\rm R}
(\eta', \eta) A^\psi_{\rm R} (\eta', \eta) = - 2^4 \,|z|^2\, D_F\,
\delta_{\eta'\eta, + } \prod_{n = 1}^\infty (1 + \lambda\, |z|^{2n})
(1 + \lambda^{- 1} \, |z|^{2n} ) ( 1 + |z|^{2n})^6, \ee for the R-R
sector. Note that we have $|z| = e^{- \pi t}$ above and in (95) we
have followed the prescription given in \cite{Billo:1998vr,Di
Vecchia:1999rh} not to separate the contributions from matter fields
$\psi^\mu$ and superghosts in the R-R sector in order to avoid the
complication due to the respective zero modes.  Also in the above,
we have \be C_F = \left\{\begin{array} {cc} \sqrt{(1 - f_1^2)(1 -
f_2^2)} & {\rm for}
\,(F, D_p), \\
&\\
  \sqrt{(1 + f_1^2)(1 + f_2^2)} &\quad\,\, {\rm for} \,(D_{p -2}, D_p),
\end{array}\right. \ee  \be D_F  = \left\{\begin{array}
{cc} \frac{1 - f_1 f_2}{\sqrt{(1 - f_1^2)(1 - f_2^2)}} & {\rm for}
\,(F, D_p), \\
&\\
  \frac{1 + f_1 f_2} {\sqrt{(1 + f_1^2)(1 + f_2^2)}} &\quad\,\, {\rm for} \,(D_{p -2}, D_p),
\end{array}\right. \ee and \be \lambda + \lambda^{- 1} =  2 (2 D_F^2 - 1) = \left\{\begin{array}
{cc} 2 \frac{(1 + f_1^2) (1 + f_2^2) - 4 f_1 f_2}{(1 - f_1^2)(1 -
f_2^2)} & {\rm for}
\,(F, D_p), \\
&\\
 2 \frac{(1 - f_1^2)(1 - f_2^2) + 4 f_1 f_2}{ (1 + f_1^2)(1 +
f_2^2)} &\quad\,\, {\rm for} \,(D_{p -2}, D_p),
\end{array}\right. \ee with the previously given \be f_i  = \left\{\begin{array}
{cc} - \frac{m_i}{\triangle^{1/2}_{(m_i, n_i)}} & {\rm for}
\,(F, D_p), \\
&\\
  - \frac{m_i}{n_i} &\quad\,\, {\rm for} \,(D_{p -2}, D_p),
\end{array}\right. \ee where $i = 1, 2$ and the explicit expression for
$\triangle_{(m_i, n_i)}$ is given in (42).

In calculating $A^X$ and $A^\psi (\eta', \eta)$ as given explicitly
above, we have made use of an important property for the S matrix
\be {S^T}_\mu\,^\rho S_\rho\,^\nu = \delta_\mu\,^\nu,\ee with $T$
denoting the transpose. We can check this using, for example, the
explicit expression (11) for $S_{\mu\nu}$ with the indices raised or
lowered using the corresponding metric. This property enables us to
perform unitary transformations of the respective operators in the
boundary states (6)-(8) such that the $S$ matrix appearing in one of
the boundary states, for example, in the boundary state originally
denoting as `1' above, completely disappears, while leaving the
other one (originally denoting as `2') with a new S matrix as $S =
S_2 S^T_1$, in the course of evaluating the respective $A^X$ or
$A^\psi$. This new S matrix shares the same property (100) as the
original $S_1$ and $S_2$ do but its determinant is always equal to
one. Therefore this S matrix under consideration can always be
diagonalized to give two eigenvalues $\lambda$ and $\lambda^{-1}$
with their sum as given in (98) above and the other eight
eigenvalues all equal to one. This is the basis for the structure
appearing in the contributions due to the respective oscillators to
the $A^X$ and $A^\psi (\eta', \eta)$ as given in (93)-(95) above.

We can now have the vacuum amplitude in the NS-NS sector as \bea
\Gamma_{\rm NS}& =& \,_{\rm NS}\langle B^1| D |B^2 \rangle_{\rm NS}
\nn &=& \frac{n_1 n_2 \, c_p^2\, V_{p + 1}\, C_F}{32 \pi (2 \pi^2
\alpha')^{\frac{7 - p}{2}}} \int_0^\infty \frac{d t}{t} \, e^{-
\frac{Y^2}{2 \pi \alpha' t}} \, t^{ - \frac{7 - p}{2}} \nn
&\,&\,\qquad\quad \times |z|^{ - 1} \left[\prod_{n = 1}^\infty
\frac{(1 + \lambda\, |z|^{2n - 1})(1 + \lambda^{- 1} \,|z|^{2n - 1})
(1 + |z|^{2n - 1})^6}{(1 - \lambda\, |z|^{2n})(1 - \lambda^{- 1} \,
|z|^{2n}) (1 - |z|^{2n})^6}\right.\nn &\,&\,\qquad\qquad\qquad\left.
- \prod_{n = 1}^\infty \frac{(1 - \lambda\, |z|^{2n - 1})(1
-\lambda^{- 1} \,|z|^{2n - 1}) (1 - |z|^{2n - 1})^6}{(1 - \lambda\,
|z|^{2n})(1 - \lambda^{- 1} \, |z|^{2n}) (1 -
|z|^{2n})^6}\right],\eea where we have used the GSO projected
boundary state in (1) for $|B^i\rangle_{\rm NS}$ (i = 1, 2) with
$B^i$ as defined previously and have made use of the matrix elements
in (93) and (94). Also we have used in the above \be \int_{|z| \le
1} \frac{d^2 z}{|z|^2} = 2\pi^2 \int_0^\infty d t, \ee with $|z| =
e^{- \pi t}$. The corresponding vacuum amplitude in the R-R sector
is now \bea \Gamma_{R} &=& \,_{\rm R}\langle B^1| D |B^2
\rangle_{\rm R} \nn &=& - \frac{n_1 n_2\, c_p^2 \, V_{p + 1}\, C_F
\,D_F}{2 \pi (2 \pi^2 \alpha')^{\frac{7 - p}{2}}} \int_0^\infty
\frac{d t}{t}\,e^{- \frac{Y^2}{2 \pi \alpha' t}} \, t^{ - \frac{7 -
p}{2}} \nn &\,& \qquad \quad \times \prod_{n = 1}^\infty \frac{(1 +
\lambda\, |z|^{2n})(1 + \lambda^{- 1} \,|z|^{2n}) (1 +
|z|^{2n})^6}{(1 - \lambda\, |z|^{2n})(1 - \lambda^{- 1} \, |z|^{2n})
(1 - |z|^{2n})^6},\eea where we have used the GSO projected boundary
state in (2) for $|B^i\rangle_{\rm R}$ ($i = 1, 2$) again with $B^i$
as defined previously  and made use of the matrix elements in (93)
and (95) as well as the equation (102). In the above, we always
assume both $n_1 $ and $n_2 $ are positive integers and the p-branes
in the non-threshold bound states are both D$_p$ branes (or both
anti D$_p$ branes). In the case when the p-branes in either of the
non-threshold bound states (but not both) are anti D$_p$ branes, the
corresponding $\Gamma_{\rm R}$ will switch sign from the one above
but the $\Gamma_{\rm NS}$ will remain the same. In what follows, we
will focus on that the p-branes in both non-threshold bound states
are D$_p$-branes, i.e., (101) and (103) are valid. The case when the
p-branes in either of the bound states are anti D$_p$-branes can be
similarly analyzed.

We would like to pause here to make a few checks of the above
results (101) and (103) against known ones. When we set $n_1 = n_2 =
1$ and switch off the worldvolume gauge fields, i.e., setting $f_1 =
f_2 = 0$ ( therefore $C_F = D_F = 1$ and $\lambda = \lambda^{-1} =
1$),  our above $\Gamma_{\rm NS}$ and $\Gamma_{\rm R}$ agree with
the well-known results between two identical Dp-branes placed
parallel to each other and separated by a distance $Y$. For example,
our results completely agree with the calculations given in Eq.
(9.285) and Eq. (9.289) in \cite{Di Vecchia:1999rh} when we set $p =
p'$, i.e., $\nu = 0$, in their case if we notice that \be
\frac{c_p^2}{32 \pi (2\pi^2 \alpha')^{\frac{7 - p}{2}}} =
\frac{1}{(8 \pi^2 \alpha')^{\frac{p + 1}{2}}}\times \frac{1}{2},\ee
where we have used the explicit expression (5) for $c_p$. For the
case of (F, D$_p$) bound state, when two such bound states are
identical, i.e., $f_1 = f_2 = - m/\triangle^{1/2}_{(m, n)}$, the
results for $\Gamma_{\rm NS}$ and $\Gamma_{\rm R}$ with the string
coupling set to unit  were given in \cite{DiVecchia:1999uf}  as
mentioned earlier. Applying the same conditions to our calculations
for the (F, D$_p$) case, we again find perfect agreements if we make
use of (104) and notice the following: 1) $S_1 = S_2$, therefore the
matrix $S = S_1 S_2^T $ is now a unit matrix and so $\lambda =
\lambda^{-1} = 1$; 2) \be D_F = 1, \qquad C_F =  1 - f^2 =
\frac{n^2}{g_s^2 \triangle_{(m,n)}}\ee with $\triangle_{(m,n)}$
given in (42) and $g_s$ set equal to unit; 3) Their integration
variable $t$ is $\pi$ times ours.

The total vacuum amplitude is now \bea \Gamma &=& \Gamma_{\rm NS} +
\Gamma_{\rm R} \nn &=& \frac{n_1 n_2 \, V_{p + 1}\, C_F}{2  (8 \pi^2
\alpha')^{\frac{1 + p}{2}}} \int_0^\infty \frac{d t}{t} \, e^{-
\frac{Y^2}{2 \pi \alpha' t}} \, t^{ - \frac{7 - p}{2}} \nn
&\,&\,\quad \times \left\{ |z|^{ - 1} \left[\prod_{n = 1}^\infty
\frac{(1 + \lambda\, |z|^{2n - 1})(1 + \lambda^{- 1} \,|z|^{2n - 1})
(1 + |z|^{2n - 1})^6}{(1 - \lambda\, |z|^{2n})(1 - \lambda^{- 1} \,
|z|^{2n}) (1 - |z|^{2n})^6}\right.\right.\nn
&\,&\,\qquad\qquad\qquad\left. - \prod_{n = 1}^\infty \frac{(1 -
\lambda\, |z|^{2n - 1})(1 -\lambda^{- 1} \,|z|^{2n - 1}) (1 -
|z|^{2n - 1})^6}{(1 - \lambda\, |z|^{2n})(1 - \lambda^{- 1} \,
|z|^{2n}) (1 - |z|^{2n})^6}\right]\nn &\,&\,\qquad\quad \left. -
2^4\, D_F\,\prod_{n = 1}^\infty \frac{(1 + \lambda\, |z|^{2n})(1 +
\lambda^{- 1} \,|z|^{2n}) (1 + |z|^{2n})^6}{(1 - \lambda\,
|z|^{2n})(1 - \lambda^{- 1} \, |z|^{2n}) (1 -
|z|^{2n})^6}\right\},\eea where we have used the explicit expression
(5) for $c_p$ and Eq. (104). This is our basic result of this paper
in addition to the long-distance one given in the previous section.
At first look, this is completely different from the calculation
given in \cite{Green:1996um} for $p = 1$, i.e. the D-string case in
the Wick rotated version using the light-cone boundary state. In
what follows, we will show that our result above is indeed the same
as theirs for $p = 1$ using various $\theta$-function relations. For
this purpose, let us express our amplitude (106) in terms of
$\theta$-functions and the Dedekind $\eta$-function with their
standard definitions as given, for example, in \cite{polbookone}. We
then have \bea \Gamma &=& \frac{n_1 n_2 \, V_{p + 1} \, C_F\, \sin
\pi \nu}{ (8 \pi^2 \alpha')^{\frac{1 + p}{2}}} \int_0^\infty \frac{d
t}{t} \, e^{- \frac{Y^2}{2 \pi \alpha' t}} \, t^{ - \frac{7 - p}{2}}
\nn &\,&\quad\times \frac{1}{\eta^9 (it)} \left[\frac{\theta_3
(\nu|it)\, \theta^3_3 (0|it)}{\theta_1 (\nu |it)} - \frac{\theta_4
(\nu | it) \theta^3_4 (0 | it)} {\theta_1 (\nu | it)} -
\frac{\theta_2 (\nu | it) \theta^3_2 (0 | it)}{\theta_1 (\nu | it)}
\right],\eea where we have defined $\lambda = e^{2\pi {\rm i} \nu}$
and used the fact $\cos\pi \nu = D_F$ which can be obtained from
$\lambda + \lambda^{-1} = 2 (2 D_F^2 - 1)$ as given in (98). Note
that $\nu =  i \nu_0$ with $0 \le \nu_0 < \infty$ for the case of
(F, D$_p$) while $\nu = \nu_0$ with $0 \le \nu_0 < 1$ for (D$_{p -
2}$, D$_p$).  Further $\nu_0 \rightarrow \infty$ when $f_1 \neq f_2$
and either of $|f_i| \rightarrow 1$ (or both $|f_i| \rightarrow 1$
when $f_1 = -f_2$) in the former case while $\nu_0 \rightarrow 1$
when $f_1 = - f_2$ with $|f_i| \rightarrow \infty$ in the latter
case but $\nu_0 = 0$ when $f_1 = f_2$ in both cases. Now we use the
following identify for $\theta$-functions \be 2 \, \theta^4_1 (\nu |
\tau) = \theta_3 (2 \nu | \tau)\, \theta^3_3 (0 | \tau) - \theta_4
(2 \nu | \tau)\, \theta^3_4 (0 | \tau) - \theta_2 (2 \nu | \tau)\,
\theta^3_2 (0 | \tau), \ee which is obtained from (iv) on page 468
in \cite{whittaker-watson}\footnote{In obtaining the above identity
from the more general one (iv) there, we have made choices of
variables $x' = y' = z' = 0$ and $w' = 2 z$ which give $w = - z$ and
$x = y = z$ in their notation. Note also that their notation for
$\theta$-functions is $\theta_r (z) = \theta_r (z | \tau)$ with $r =
1, 2, 3, 4$. We also use the facts that $\theta_1 (0 | \tau) = 0$
and $\theta_1 ( - z |\tau) = - \theta_1 (z | \tau)$.}. With the
identity (108), the amplitude (107) is greatly simplified to \bea
\Gamma &=& \frac{2\, n_1 n_2 \, V_{p + 1} \, C_F\, \sin \pi \nu}{ (8
\pi^2 \alpha')^{\frac{1 + p}{2}}} \int_0^\infty \frac{d t}{t} \,
e^{- \frac{Y^2}{2 \pi \alpha' t}} \, t^{ - \frac{7 - p}{2}}
\frac{1}{\eta^9 (it)} \frac{\theta^4_1 (\frac{\nu}{2} |
it)}{\theta_1 (\nu |it)},\nn &=& \frac{U(m_1, n_1; m_2, n_2)\, V_{p
+ 1} }{ 2 (8 \pi^2 \alpha')^{\frac{1 + p}{2}}} \frac{\sin \pi
\nu}{\sin^4 \frac{\pi \nu}{2}}\,  \int_0^\infty \frac{d t}{t} \,
e^{- \frac{Y^2}{2 \pi \alpha' t}} \, t^{ - \frac{7 - p}{2}}
\frac{1}{\eta^9 (it)} \frac{\theta^4_1 (\frac{\nu}{2} |
it)}{\theta_1 (\nu |it)},\nn &=& \frac{4\,U(m_1, n_1; m_2, n_2)\,
V_{p + 1} }{ (8 \pi^2 \alpha')^{\frac{1 + p}{2}}} \int_0^\infty
\frac{d t}{t} \, e^{- \frac{Y^2}{2 \pi \alpha' t}} \, t^{ - \frac{7
- p}{2}} \nn &\,&\qquad\qquad \qquad\qquad\quad\times \prod_{n =
1}^\infty \frac{\left(1 - e^{i\pi \nu} |z|^{2n}\right)^4 \left(1 -
e^{- i \pi \nu} |z|^{2n}\right)^4}{\left(1 - |z|^{2n}\right)^6 \,
\left(1 - e^{2 i \pi \nu} |z|^{2n} \right)\, \left(1 - e^{- 2 i \pi
\nu} |z|^{2n}\right)},\eea where in the second equality, we have
made use of \be \sin^4 \frac{\pi \nu}{2} = \frac{1}{4} \left(\cos
\pi \nu - 1\right)^2 = \frac{1}{4} (D_F - 1)^2, \qquad n_1 n_2 C_F
(D_F - 1)^2 = U (m_1, n_1; m_2, n_2),\ee with $U (m_1, n_1; m_2,
n_2) = U_{\rm NS} (m_1, n_1; m_2, n_2) - U_{\rm R} (m_1, n_1; m_2,
n_2)$ as given by (84) for either case under consideration and with
the respective quantization for $f_i$ as given previously, and in
the third equality we have made use of explicit expressions for the
Dedekind $\eta$-function and the theta-function $\theta_1$.

One can check now that our above amplitude in the present various
forms does agree with the calculations given in \cite{Green:1996um}
for the $p = 1$ case in the light-cone approach up to an overall
constant factor\footnote{In making the comparison, we have
considered both zero-mode contribution (77) and the oscillator
contribution (82) in \cite{Green:1996um} for the magnetic flux. For
the case of electric flux, one should send $f_1 \rightarrow i f_1$
and $f_2 \rightarrow i f_2$ as well as $\alpha \rightarrow i \alpha$
as mentioned there. In their calculation, the volume factor was not
considered  and the overall constant factor difference mentioned in
the text should not be concerned here since it is well-known that
the light-cone calculations alone cannot fix the overall constant.}
of $1/(8 \pi^6)$. In making the comparison, we need also to consider
that in their calculations they chose $\alpha' = 2$ and their
parameter $\alpha$ is related to our $\nu$ as $\alpha = 2 \pi \nu$.

We now consider the large $Y$ limit of the amplitude (109). This
amounts to accounting for the massless-mode contribution of closed
string and therefore the result should agree with our low-energy
effective field theory calculations performed in the previous
section. We will find that this is indeed true\footnote{One can also
show that $\Gamma_{\rm NS}$ (101) and $\Gamma_{\rm R}$ (103) give
also their corresponding low energy limits as discussed in the
previous section in a similar fashion.}. For large $Y$, the
separation dependent exponential suppression factor in (109) implies
that the contribution to the amplitude comes from the large t
integration. Note that for large t, $|z| = e^{- \pi t} \rightarrow
0$ and \be \theta_1 (\nu | it) \rightarrow 2 e^{- \frac{\pi t}{4}}
\sin \pi\nu, \qquad \theta_1 (\frac{\nu}{2} | it) \rightarrow 2 e^{-
\frac{\pi t}{4}} \sin \frac{\pi\nu}{2}, \qquad \eta (i t)
\rightarrow e^{- \frac{\pi t}{12}}.\ee So \bea \Gamma &\rightarrow&
\frac{ U(m_1, n_1; m_2, n_2) \, V_{p + 1}}{2 (8 \pi^2
\alpha')^{\frac{1 + p}{2}}} \frac{\sin \pi\nu}{\sin^4 \frac{\pi
\nu}{2}} \int_0^\infty \frac{d t}{t} \, e^{- \frac{Y^2}{2 \pi
\alpha' t}} \, t^{ - \frac{7 - p}{2}} \frac{1}{e^{- \frac{3\pi
t}{4}}} \frac{ 2^4\, e^{- \pi t}\, \sin^4 {\frac{\pi\nu}{2}}}{2\,
e^{- \frac{\pi t}{4}} \sin\pi \nu},\nn &=\,& \frac{4 \, U(m_1, n_1;
m_2, n_2) \, V_{p + 1} }{ (8 \pi^2 \alpha')^{\frac{1 + p}{2}}}
\int_0^\infty \frac{d t}{t} \, e^{- \frac{Y^2}{2 \pi \alpha' t}} \,
t^{ - \frac{7 - p}{2}}, \nn &=\,& \frac{4\, U(m_1, n_1; m_2, n_2)\,
V_{p + 1} }{ (8 \pi^2 \alpha')^{\frac{1 + p}{2}}}
\left(\frac{2\pi\alpha'}{Y^2}\right)^{\frac{7 - p}{2}} \Gamma
\left(\frac{7 - p}{2}\right), \nn &=\,& \frac{C(m_1, n_1; m_2,
n_2)}{Y^{7 - p}},\eea where $C (m_1, n_1; m_2, n_2)$ is given by
(83). So this is in complete agreement with our low-energy result
(82), as expected, which in turn shows that even our normalization
constant is also correct. In reaching the last equality, we have
made use of (104) and $(7 - p)\Omega_{8 - p} = 4 \pi \pi^{(7 -
p)/2}/\Gamma ((7 - p)/2)$ with $\Omega_q$ the volume of unit
q-sphere.

 The interaction amplitude (109) vanishes when $U(m_1, n_1; m_2, n_2) =
 0$ which gives $m_1/n_1 = m_2/n_2$ (note $n_1 n_2 > 0$)
 as shown in the previous section (now  $\nu = 0$ since $f_1 = f_2$),
 reflecting the BPS property of the
system.  If we take one pair of integers, say the pair $(m_2, n_2)$,
as co-prime, then the vanishing amplitude would need $(m_1, n_1) = k
(m_2, n_2)$ with $k$ a positive integer. Note that unlike the single
brane case, the non-threshold bound states have infinite many stable
fundamental states with each characterized by a different pair of
co-prime integers $(m, n)$. When placing a brane with a pair of
integers $k (m, n)$ parallel to one with its pair of integers $k'
(m, n)$, we have the system breaking no supersymmetry and being BPS
if $k k' > 0$, i.e., integer $k$ and integer $k'$ have the same
sign. When $(m_1, n_1)$ and $(m_2, n_2)$ are both co-prime, the
interaction vanishes only if $(m_1, n_1) = (m_2, n_2)$. Further when
none of the above is satisfied, we have $U (m_1, n_1; m_2, n_2) >
0$. Note that each numerator in the infinite product in the
integrand of (109) \be \left(1 - e^{i\pi \nu} |z|^{2n}\right)^4
\left(1 - e^{- i\pi \nu} |z|^{2n}\right)^4 = \left(1 - 2
\cos\pi\nu\, |z|^{2n} + |z|^{4n}\right)^4 > 0, \ee so the sign of
the interaction amplitude will depend on that of the factor in each
denominator in the infinite product in the integrand \be \left(1 -
e^{2 i \pi \nu} |z|^{2n}\right) \left(1 - e^{- 2 i\pi \nu}
|z|^{2n}\right)= \left(1 - 2 \cos{2\pi\nu}\, |z|^{2n} +
|z|^{4n}\right), \ee which is always positive for the case of (D$_{p
- 2}$, D$_p$) (now $\nu$ is real) while it is positive for large $t$
but it can be negative for small $t$ for the case of (F, D$_p$) for
which $\nu$ is purely imaginary. So for the case of (D$_{p - 2}$,
D$_p$), the interaction amplitude is now greater than zero and is
solely determined by the positiveness of $U (m_1, n_1; m_2, n_2)$.
In this aspect it shares the same feature as its long distance
interaction shown in the previous section, reflecting the attractive
nature of the interaction. For the case of (F, D$_p$), while the
long distance interaction amplitude is again now greater than zero
(implying attractive interaction ) and is also solely determined by
the positiveness of the corresponding $U (m_1, n_1; m_2, n_2)$ as
shown in the previous section, the sign of the small separation
amplitude (corresponding to small $t$ contribution) is uncertain in
the present representation of integration variable $t$ since even
with the factor in (114) less than zero, the sign of  the product of
infinite such factors in the integrand remains indefinite. So one
would expect some interesting physics to appear in this case for
small $t$.

The small $t$ contribution to the amplitude mainly concerns about
the physics for small separation $Y$. The appropriate frame for
describing the underlying physics as well as the analytic structure
as a function of the separation in the short cylinder limit $t
\rightarrow 0$ is in terms of an annulus, which can be achieved by
the Jacobi transformation $t \rightarrow t' = 1/t$. This is also
stressed in \cite{Douglas:1996yp} that the lightest open string
modes now contribute most and the open string description is most
relevant. So in terms of the annulus variable $t'$, noting \bea \eta
(\tau) &=& \frac{1}{\left(- i \tau\right)^{1/2}}\, \eta \left(-
\frac{1}{\tau}\right),\nn \theta_1 (\nu|\tau) &=&  i \,\frac{\,e^{-
i \pi \nu^2/\tau}}{\left(- i \tau\right)^{1/2}} \,\theta_1
\left(\left.{\frac{\nu}{\tau}}\right|- \frac{1}{\tau}\right),\eea
the second equality in (109) now becomes \bea \Gamma &=& - i\,
\frac{U(m_1, n_1; m_2, n_2)\, V_{p + 1} }{ 2 (8 \pi^2
\alpha')^{\frac{1 + p}{2}}} \frac{\sin \pi \nu}{\sin^4 \frac{\pi
\nu}{2}}\,  \int_0^\infty \frac{d t'}{t'} \, e^{- \frac{Y^2 t'}{2
\pi \alpha' }} \, t'^{ \frac{1 - p}{2}} \frac{1}{\eta^9 (it')}
\frac{\theta^4_1 (\frac{- i\nu t'}{2} | it')}{\theta_1 (-i \nu
t'|it')},\nn &=& - i\, \frac{4 \,U(m_1, n_1; m_2, n_2)\, V_{p + 1}
}{ (8 \pi^2 \alpha')^{\frac{1 + p}{2}}} \frac{\sin \pi \nu}{\sin^4
\frac{\pi \nu}{2}}\, \int_0^\infty \frac{d t'}{t'} \, e^{- \frac{Y^2
t'}{2 \pi \alpha' }} \, t'^{ \frac{1 - p}{2}} \frac{\sin^4
\left(\frac{- i \pi \nu t'}{2}\right)}{\sin \left(- i \pi \nu
t'\right)} \nn &\,& \qquad\qquad \quad\times \prod_{n = 1}^\infty
\frac{\left(1 - e^{\pi \nu t'} |z|^{2n}\right)^4 \left(1 - e^{-\pi
\nu t'} |z|^{2n}\right)^4}{\left(1 - |z|^{2n}\right)^6 \, \left(1 -
e^{2 \pi \nu t'} |z|^{2n} \right)\, \left(1 - e^{- 2  \pi \nu t'}
|z|^{2n}\right)},\eea with now $|z| = e^{- \pi t'}$. We  follow
\cite{Green:1996um} to discuss the underlying analytic structure and
the possible associated physics of the amplitude of (116). For the
case of (D$_{p - 2}$, D$_p$), we limit ourselves to the interesting
non-BPS amplitude, i.e., $\nu = \nu_0$ with $0 < \nu_0 < 1$, and for
this the above amplitude is real and has no singularities unless $Y
\leq 2\pi \sqrt{\nu \alpha'}$, i.e. on the order of string scale,
for which the integrand is dominated by, in the short cylinder limit
$t' \rightarrow \infty$, \be \, \lim_{t' \rightarrow \infty}
\frac{e^{- \frac{Y^2 t'}{2\pi \alpha'}} \theta^4_1 (- i \pi \nu t'/2
|i t')}{i\,\eta (it') \theta_1 (- i \pi \nu t'| it')} \sim \lim_{t'
\rightarrow \infty} \frac{e^{- \frac{Y^2 t'}{2\pi \alpha'}} \sin^4(-
i \pi \nu t'/2)}{i\,\sin (- i \pi\nu t')} \sim \lim_{t'\rightarrow
\infty} e^{- \frac{t'}{2\pi\alpha'}(Y^2 - 2\pi^2 \nu \alpha')}. \ee
The contribution of the annulus to the vacuum amplitude (free
energy) should be real if the integrand in (116) have no simple
poles on the positive $t'$-axis  since the imaginary part of the
amplitude is given by the sum of residues at the poles times $\pi$
due to the integration contour passing to the right of all poles as
dictated by the proper definition of the Feynman
propagator\cite{Bachas:1992bh}. In the present case, the amplitude
appears purely real but there are no simple poles on the positive
$t'$-axis, therefore giving zero imaginary amplitude, i.e., zero
pair-production (absorptive) rate, which is consistent with the
conclusion reached in \cite{Schwinger:1951nm} in quantum field
theory context and also pointed out in a similar context in
\cite{Bachas:1992zr}. When $Y \leq \pi \sqrt{2\nu_0 \alpha'}$, i.e.,
on the order of string scale, the integration in (116) diverges and
this therefore gives  a divergent amplitude which indicates the
breakdown of the calculations and behaves similarly to the situation
of brane/antibrane systems as studied in \cite{Banks:1995ch,
Lu:2007kv}, signalling the possible onset of tachyonic instability
now caused instead by the magnetic fluxes\footnote{Without the
presence of the magnetic flux, the system is a BPS one and the
amplitude vanishes. With the presence of the magnetic flux, in
addition to the evidence given in the text, that the open string
tachyon mode appears to arise is also indicated from the leading
term $e^{\pi \nu t'}$, which diverges in the short cylinder limit
$t' \rightarrow \infty$, in the expansion of the $\theta$-functions
and $\eta$-function in (116) in the open string channel. } and the
relaxation of the system to form new non-threshold bound state.
However,  the detail of this requires further dynamical
understanding.

Let us move to the case of (F, D$_p$). We have now $\nu = i \nu_0$
with $0 < \nu_0 < \infty$ ($\nu_0 = 0$ corresponds to BPS case and
is not considered here). The amplitude (116) is now \bea \Gamma  &=&
 \frac{4 \,U(m_1, n_1; m_2, n_2)\, V_{p + 1} }{ (8 \pi^2
\alpha')^{\frac{1 + p}{2}}} \frac{\sinh \pi \nu_0}{\sinh^4 \frac{\pi
\nu_0}{2}}\, \int_0^\infty \frac{d t'}{t'} \, e^{- \frac{Y^2 t'}{2
\pi \alpha' }} \, t'^{ \frac{1 - p}{2}} \frac{\sin^4 \left(\frac{
\pi \nu_0 t'}{2}\right)}{\sin \left( \pi \nu_0 t'\right)} \nn &\,&
\qquad\qquad \quad\times \prod_{n = 1}^\infty \frac{\left(1 - e^{i
\pi \nu_0 t'} |z|^{2n}\right)^4 \left(1 - e^{- i\pi \nu_0 t'}
|z|^{2n}\right)^4}{\left(1 - |z|^{2n}\right)^6 \, \left(1 - e^{ 2 i
\pi \nu_0 t'} |z|^{2n} \right)\, \left(1 - e^{- 2 i \pi \nu_0
t'}|z|^{2n}\right)}.\eea Exactly the same as the $p = 1$ case given
in \cite{Green:1996um}, the above integrand has also an infinite
number of simple poles on the positive real $t'$-axis at $t' = (2 k
+ 1)/\nu_0$ with $k = 0, 1, 2, \cdots$. This leads to an imaginary
part of the amplitude, which is given as the sum over the residues
of the poles as described in \cite{Bachas:1992bh, Bachas:1995kx}.
Therefore the rate of pair production of open strings per unit
worldvolume in a constant electric flux in the present context  is
\bea {\cal W} &\equiv& - \frac{2 {\rm Im} \Gamma}{V_{p + 1}},\nn &=&
\frac{8 \,U(m_1, n_1; m_2, n_2) }{\nu_0 (8 \pi^2 \alpha')^{\frac{1 +
p}{2}}} \frac{\sinh \pi \nu_0}{\sinh^4 \frac{\pi \nu_0}{2}}\,\sum_{k
= 0}^\infty \left(\frac{\nu_0}{2 k + 1}\right)^{\frac{1 + p}{2}}
e^{- \frac{(2 k + 1) Y^2 }{2 \pi \nu_0 \alpha' }} \, \prod_{n =
1}^\infty \left(\frac{1 + e^{- 2n\pi (2 k + 1)/\nu_0}}{ 1 - e^{- 2 n
(2 k + 1)\pi/\nu_0 }}\right)^8, \nn &=& \frac{32 n_1 n_2
\left|\frac{m_1}{\triangle^{1/2}_{(m_1, n_1)}} -
\frac{m_2}{\triangle^{1/2}_{(m_2, n_2)}}\right| }{\nu_0 (8 \pi^2
\alpha')^{\frac{1 + p}{2}}}\,\sum_{k = 0}^\infty
\left(\frac{\nu_0}{2 k + 1}\right)^{\frac{1 + p}{2}} e^{- \frac{(2 k
+ 1) Y^2 }{2 \pi \nu_0 \alpha' }} \, \prod_{n = 1}^\infty
\left(\frac{1 + e^{- 2n (2 k + 1)\pi/\nu_0}}{ 1 - e^{- 2 n (2 k +
1)\pi/\nu_0 }}\right)^8,\nn\eea where $\triangle_{(m, n)}$ is
defined in (42) and $\nu_0$ can be determined from \be \cosh \pi
\nu_0 = \frac{g^2_s \left(\Omega - m_1 m_2\right)}{n_1 n_2} \ee with
$\Omega$ defined in (50). This rate has been calculated in different
context before \cite{Bachas:1992bh, Bachas:1995kx, Cho:2005aj,
Chen:2008jx} but as stressed in \cite{Green:1996um} for the $p = 1$
case, the rather complicated sum over spin structures obtained in
those papers reduces to our simple expression of (116) or (118) or
(119). Note that the above rate is suppressed by the brane
separation and the integer $k$ but increases with the value of
$\nu_0$ which is expected. Let us consider $\nu_0 \rightarrow 0$ and
$\nu_0 \rightarrow \infty$ limits for the above rate, respectively.
The former limit corresponds to the near extremal limit for which we
can set $f_1 = f_2 + \epsilon$ with $|\epsilon| \ll 1$ while the
latter corresponds to the critical field limit for which one can set
either $|f_i| \rightarrow 1$ while keeping the other less than unit
( but fixed) or set $f_1 = - f_2 $ with both $|f_i| \rightarrow 1$
as mentioned earlier. The definition for $f_i$ with $i = 1, 2$ is
given in (99). For the near extremal limit, we have, to leading
order, \be \nu_0 \approx \frac{|\epsilon|}{\pi (1 - f_2^2)},\ee the
rate (119) is now well approximated by the $k = 0$ term as \be {\cal
W} \approx \frac{32 n_1 n_2\,|\epsilon|}{(8\pi^2 \alpha')^{\frac{1 +
p}{2}}} \left(\frac{|\epsilon|}{\pi (1 - f_2^2)}\right)^{\frac{p -
1}{2}} e^{- \frac{Y^2  (1 - f_2^2)}{2\alpha' |\epsilon|}}, \ee very
tiny as expected. For the critical field limit mentioned above, now
$\nu_0 \rightarrow \infty$ and it is easy to see that each term in
the summation of (119) diverges and so does the rate, signalling
also an instability as mentioned in a similar context in
\cite{Porrati:1993qd}.

\section{Summary}
In this paper, we calculate explicitly the interaction amplitude
between two ($F, D_p$) or ($D_{(p - 2)}, D_p$) non-threshold bound
states with a separation. In doing so, we make use of their
respective boundary state representation with a quantized
world-volume electric (or magnetic) flux. Each such non-threshold
bound state is therefore characterized by a pair of integers $(m_i,
n_i)$ with $i = 1, 2$. When the two bound states are (D$_{p - 2}$,
D$_p$), the interaction is in general attractive but this remains so
and can be certain only at large brane separation when the two
states are (F, D$_p$). In both cases, the interaction vanishes only
if $m_1/n_1 = m_2/n_2$ and  $n_1 n_2
> 0$. We also calculate the respective long-distance
interaction independently from the low energy field theory approach
and each agrees with the long-distance part of the corresponding
general string amplitude. We also study the analytic structure of
the amplitude and in particular we calculate the rate of pair
production of open strings for the case of (F, D$_p$). In general,
one expects that the interacting system is unstable and will relax
itself by releasing the exceed energy via so-called tachyonic
condensation\cite{polbooktwo} to form eventually a BPS non-threshold
bound state, characterized by a pair of integers ($m_1 + m_2, n_1 +
n_2$). If $m_1 + m_2$ and $n_1 + n_2$ are co-prime, this state will
be stable otherwise it will be marginally unstable. Similar to the
brane/antibrane systems studied in \cite{Banks:1995ch,Lu:2007kv},
the open string tachyonic condensation manifests itself for the case
of (D$_{p - 2}$, D$_p$) by showing a divergent amplitude but now
caused by the presence of magnetic fluxes when the brane separation
is on the order of string scale. However, for the case of (F,
D$_p$), this manifests itself by the pair production of open strings
which takes the exceed energy away so that the system can lower its
energy and relax itself to form the final BPS bound state. By all
means, what has been said here is just an indication being
responsible for forming the final BPS states of the systems under
consideration. To determine whether this actually leads to the
formation of final BPS states requires  a more detailed dynamical
understanding which is beyond the scope of this paper.

\vspace{.5cm}

\noindent {\bf Acknowledgements}
 \vspace{2pt}

 JXL would like to thank Zhi-Zhong Xing and other members of the Theory Division
 at the CAS Institute of High Energy Physics, the TPCSF-CAS and the KITPC-CAS at Beijing
 for their hospitality. We would  like to thank Dao-Neng Gao and Huan-Xiong Yang for useful
 discussion,  Zhao-Long Wang and Rong-Jun Wu for reading the manuscript and pointing out a few typos.
 We acknowledge support by grants from the Chinese Academy of
Sciences, a grant from 973 Program with grant No: 2007CB815401 and
grants from the NSF of China with Grant No:10588503 and 10535060.

\end{document}